\documentstyle[floats,prd,aps,tighten]{revtex}
\input{epsf.sty}
\newcommand{\be}{\begin{equation}}
\newcommand{\ee}{\end{equation}}
\newcommand{\bea}{\begin{eqnarray}}

\newcommand{\eea}{\end{eqnarray}}
 
\newcommand{\ba}{\begin{array}}
\newcommand{\ea}{\end{array}}

\def\ls{\mathrel{\lower4pt\vbox{\lineskip=0pt\baselineskip=0pt\hbox{$<$}\hbox{$\sim$}}}}\def\gs{\mathrel{\lower4pt\vbox{\lineskip=0pt\baselineskip=0pt
\hbox{$>$}\hbox{$\sim$}}}}
\begin{document}
%\setlength{\topmargin}{+0.7 truecm}
%@twocolumnfalse\endcsname

\title{Cosmic optical activity in the spacetime of a 
scalar-tensor screwed cosmic string}

\author{V. B. Bezerra$^1$\thanks{valdir@fisica.ufpb.br}, 
H. J. Mosquera Cuesta$^2$\thanks{hermanjc@cbpf.br} and C.N. Ferreira$^{2,3,4}$\thanks{crisnfer@cbpf.br}} 

\address{$^1$Departamento de F\'{\i}sica, Universidade Federal da Para\'{\i}ba, 
58059-970, Jo\~ao Pessoa, PB, Brazil\\$^2$Centro Brasileiro de Pesquisas F\'{\i}sicas, 
Rua Dr. Xavier Sigaud 150, Urca 22290-180, Rio de Janeiro, RJ, Brazil\\$^3$Instituto de 
F\'{\i}sica, Universidade Federeal do Rio de Janeiro
Caixa postal 68528, 21945-910, Rio de Janeiro, RJ, Brazil\\$^4$Grupo de F\'{\i}sica 
Te\'orica Jos\'e Leite Lopes, Petr\'opolis, RJ, Brazil
}

\date{\today}
\maketitle

\begin{abstract}
Measurements of radio emission from distant galaxies and quasars verify that 
the polarization vectors of these radiations are not randomly oriented as 
naturally expected. This peculiar phenomenon suggests that the spacetime 
intervening between the source and observer may be exhibiting some sort of 
optical activity, the origin of which is not known. In the present paper we 
provide a plausible explanation to this phenomenon by  investigating the 
r\^ole played by a Chern-Simons-like term in the  background of an ordinary 
or superconducting screwed cosmic string in a scalar-tensor gravity. 
We discuss the possibility that the excess in polarization of the light 
from radio-galaxies and quasars can be understood as if the electromagnetic 
waves emitted by these cosmic objects interact with a scalar-tensor screwed 
cosmic string through a Chern-Simons coupling. We use current astronomical 
data to constrain possible values for the coupling constant of this theory, 
and show that it turns out to be: $\lambda \sim 10^{-26}$~eV, which is two 
orders of magnitude larger than in string-inspired theories.

\end{abstract}

%\newpage

\def\be{\begin{equation}}
\def\ee{\end{equation}}

%\newpage

%\vspace{.5 true cm}

]

\section{Introduction}

The Cosmological Principle postulates that the Universe is homogenuous and 
isotropic on large scales. This means that there exists no preferred 
direction on the sky, and that in any patch of the sky we look into we may 
expect to find roughly the same distribution of matter and radiation. This 
premise was verified  by the COBE satellite up to about $10^{-6}$ in 
temperature anisotropy in the cosmic microwave background radiation, a 
residual from the Big Bang\cite{rocky02}. However, recent measurements of 
radio and optical emission from distant radio-galaxies and quasars provides 
clear evidence that the linearly polarized light emitted by these objects 
presents an additional rotation of its polarization plane, which remains 
even after Faraday rotation, the one produced by its interaction with the 
intergalactic plasma, is extracted. This may represent evidence for 
cosmological anisotropy on large scale\cite{HL00,HL01}.

In particular, Hutsem\'ekers and Lamy \cite{HL01} found that the 
polarization vectors of light from quasars are not randomly oriented 
on the sky as standard understanding suggested. To confirm this effect 
they studied a sample of 170 optically polarized quasars with 
accurate polarization measurements. Their analysis showed 
that in some regions the polarization position angles 
appear concentrated around preferential directions, 
what suggests the existence of large-scale coherent orientations
or alignments of the quasar's radiation polarization vectors. In their 
measurements, the hypothesis of uniform distribution of polarization 
position angles may be rejected at the 1.8 $\% $ significance 
level\cite{HL01}. Though the sample seemed to be statistically not too 
much significant, further surveys  confirm its unexpected nature\cite{das00}. 
The occurrence of coherent orientations over cosmic distances, they  
claimed, seems to point towards the existence of new non-standard effects 
relevant to cosmology\cite{HL01}.

Other authors had claimed to find evidence for cosmological 
birefringence \cite{NR97,nr97-2,jain99}. This claim, initially 
contentious\cite{Eisenstein,Carroll,leahy}, was confirmed by subsequent 
studies that demonstrated the existence of such an effect up to a high 
confidence level\cite{HL00,HL01,das00,jain99}. Thenceforth, these actual 
evidences\cite{HL01,das00,jain99,kar00} give new motivations to 
investigate more detailed this possibility.

The implications of the possible existence of a preferred direction over 
cosmological distances have been discussed in the context of theories of 
gravitation\cite{will93} and observational cosmology\cite{birch82}. Such a 
phenomenon, if it exists, would imply the violation of the Lorentz 
invariance\cite{will93}, bringing unpredictable consequences for fundamental 
physics\cite{grillo00}.

The idea that the intergalactic space is a birefringent medium has been 
considered for a long time. Several potential sources of optical activity 
have already been studied. They include the scattering produced by atoms, 
by a neutrino sea, and also by a background Kalb-Ramond torsion (axion) 
field in the context of a heterotic string theory\cite{kar00}. In particular, 
the scattering-inspired theories suggested that the intergalatic medium 
contains neutral atoms and microwave radiation immersed in a neutrino or 
antineutrino sea\cite{Weinberg}. The sea is very hard to detect 
experimentally\cite{Karl,Novikov} as a result of its low energy and its 
exclusively weak interactions. Electromagnetic radiation travelling though 
the intergalatic medium interact with its components, and if this radiation 
is initially plane polarized, the plane of polarization would rotate.

In this paper, instead of the idea of a neutrino or axion-like sea 
interacting with the electromagnetic radiation or the model inspired on 
heterotic strings, we will consider that a screwed cosmic string (SCS) 
on the background, whose gravitational effects are described by a 
scalar-tensor theory, has the same effect, i. e., the plane polarized 
electromagnetic radiation from high redshifts cosmic sources has the plane 
of polarization rotated when it is travelling through the spacetime 
generated by this screwed cosmic string. In the literature, the association 
of such anisotropy with a torsion background has been 
considered\cite{kuhne97,DM97,capo99,Sengupta}. 

On the other hand, the assumption that gravity may be intermediated by a 
scalar field (or, more generally, by many scalar fields) in addition to the
usual symmetric rank-2 tensor of Einstein's general relativity has 
considerably revived in recent years\cite{will94}. It has been argued that 
gravity may be described by a scalar-tensor gravitational field, at least 
at sufficiently high energy scales\cite{Hehl,Palle,Kim}. 

From the theoretical point of view, scalar-tensor theories of gravitation, 
in which the gravitational interaction is mediated by one or several
long-range scalar fields in addition to the usual tensor field present in
Einstein's theory, are the most natural alternatives to general relativity.
In these theories the gravitational interaction is mediated by a
(spin-2) graviton and by a (spin-0) scalar field \cite{Shapiro1,Gaspperini}.
If gravity is essentially a scalar-tensor theory, there will
be direct implications for cosmology and experimental tests of the
gravitational interaction\cite{will93,will94,dam}. In particular,
any gravitational phenomena will be affected by the variation of
the gravitational {\it constant} $\tilde{G}_{0}$. At sufficiently
high energy scales where gravity becomes scalar-tensor in nature
\cite{green}, it seems worthwhile to analyze the behaviour of matter and 
radiation fields in the presence of a scalar-tensor gravitational field,
specially those generated during the early universe by objects such 
cosmic strings.  In this context, some authors have studied
solutions for cosmic strings and domain walls in Brans-Dicke
\cite{rom}, in dilaton theory \cite{greg} and in situations with 
more general scalar-tensor couplings \cite{mexg}.

The dilaton-torsion identification was previously made in a modified 
scalar-tensor theory, where the torsion field is generated by a scalar 
field\cite{Kim}. Torsion is important from the phenomenological point of 
view and it may be relevant to cosmology. This importance is associated with 
the modifications of kinematic quantities like shear, vorticity, 
acceleration, expansion and their evolution equations derived in the 
presence of torsion\cite{Palle,Trautman,Stewart,Demianski,Ellis}.

A cosmic string \cite{Vilenkin} is a topological defect that may has
been formed during phase transitions in the realm of the early
Universe\cite{Kibble}. Its gravitational field, in the context of General
Relativity is quite remarkable: a particle placed at rest around a straight, 
infinite, static cosmic string will not be attracted to it. The richness of 
new ideas this defect brought along with general relativity seems to justify
the interest in the study of this structure, and specifically the role 
played by it in the framework of cosmology due to the fact that  
it carries a large energy density and for this reason it could be a potential 
source for primordial density perturbations\cite{brandenb95}. 

In this paper we will consider the coupling of a Chern-Simons-like(CS) 
term with the spacetime of a screwed cosmic string, in a scalar-tensor 
gravity, to introduce a novel explanation of the observations suggesting 
an optical activity of spacetime. This optical activity manifests itself 
through the peculiar orientation of the polarization vectors of radio waves 
emitted by distant quasar and galaxies, which seems to indicate some kind of 
anisotropy over cosmological distances. 

This paper is organized as follows: In Section II we introduce the physics of
the scalar-tensor screwed cosmic string (SCS). In Section III, the dilatonic 
solution for the SCS is used as a background to study, in Section IV, the 
coupling of a gauge invariant Chern-Simons-like term. In Section V, 
we study the superconducting string case and a comparison is done with 
general relativity effects. In Section VI an overall discussion of our 
results is given. We stress the possibility of using this approach to 
explain the data evidencing the optical activity of the spacetime 
intervening between us and distant quasars and galaxies. Finally, in 
Section VII we provide some closing remarks.

\section{Screwed cosmic string in scalar-tensor gravity  }

The scalar-tensor theory of gravity with torsion is an extension of
Einstein's General Relativity to which a scalar field is coupled minimally 
to the gravitational field and a dynamical torsion term is considered
additionally. The action describing this coupling, here presented in the 
Jordan-Fierz frame, takes the form \cite{Kim,Gaspperini}

\begin{equation}
I=\frac 1{16\pi }\int d^4x\sqrt{{-\tilde g }}
\left[ {\tilde \phi}{ \tilde R }-\frac{\omega(\tilde \phi) }{\tilde  \phi} 
\partial _\mu \tilde  \phi\partial^\mu \tilde \phi \right] + 
I_m ( \tilde g_{\mu \nu}, \Psi)\label{acao1} \; .
\end{equation}
where $I_{m}( \tilde g_{\mu\nu}, \Psi)$ is the action of the matter,
which in the general case takes into account all fields. Here we will 
consider the presence of spinor fields in the action. The function 
$ \omega $ in a general scalar-tensor theory has a $\tilde \phi $ dependence, 
but in the specific case of the Brans-Dicke theory it is a constant. In this 
case the scalar curvature $\tilde R$, appearing in Eq.(\ref{acao1}) in the 
Jordan-Fierz frame, can be written as \cite{Kim,Gaspperini}

\begin{equation}
\tilde R = \tilde R(\{\}) + 
\epsilon \frac{\partial_{\mu} \tilde \phi \partial^{\mu} \tilde \phi}
{\tilde \phi^2} \; ,
\end{equation} 
where $\tilde R(\{\})$ is the Riemann scalar curvature in the Jordan-Fierz 
frame and $\epsilon $ is the torsion coupling constant\cite{Gaspperini}. 
It is worth to stress that in the scalar curvature $\tilde R $, the scalar
function $\tilde \phi $ (the dilaton field) can act as a source of the 
torsion field. Therefore, in the absence of string spin, the torsion field 
may be generated by the gradient of this scalar field \cite{Kim}. In this 
case the torsion can be propagated with the scalar field, and can be 
written as

\begin{equation}
S_{\mu\nu}^{\hspace{.2 true cm} \lambda} = 
( \delta^{\lambda}_{\mu} \partial_{\nu} \tilde \phi - 
\delta^{\lambda}_{\nu} \partial_{\mu}\tilde \phi)/ 
2\tilde \phi. \label{torsion}
\end{equation}

The most general affine connection $\Gamma_{\lambda \nu}^{\hspace{.3 true
cm} \alpha}$ in this theory receives a contribution from the contortion tensor
$K_{\lambda \nu}^{\hspace{.3 true cm}\alpha}$ throughout the definition

\begin{equation}
\Gamma_{\lambda \nu}^{\hspace{.3 true cm} \alpha} =
\{^{\alpha}_{\lambda \nu}\} +
 K_{\lambda \nu}^{\hspace{.3 true cm}\alpha} \; , \label{kont1}
\end{equation}
where the quantity  $ \{^{\alpha}_{\lambda \nu}\}$ is the Christoffel 
symbol computed from the metric tensor $g_{\mu \nu}$, whilst the contortion 
tensor $K_{\lambda \nu}^{\hspace{.3 true cm}\alpha}$ can be written
in  terms of the torsion field as

\begin{equation}
K_{\lambda \nu}^{\hspace{.3 true cm}\alpha} =
-\frac{1}{2}(S_{\lambda \hspace{.2 true cm} \nu}^{\hspace{.1 true
cm} \alpha} + S_{\nu \hspace{.2 true cm} \lambda}^{\hspace{.1 true
cm} \alpha} - S_{\lambda \nu}^{\hspace{.3 true cm} \alpha}) \; .
\end{equation}

Although the action proposed in Eq.(\ref{acao1}) shows explicitly this 
scalar-tensor gravity feature, for technical reasons, we will adopt the 
Einstein (conformal) frame in which the kinematic terms of the scalar 
and tensor fields do not mix. In this frame the action is given as 

\begin{equation}
{I} = \frac{1}{16\pi G} \int d^4x \sqrt{-g} \left[ R -
2g^{\mu\nu}\partial_{\mu}\phi\partial_{\nu}\phi \right]
+ I_{m}[\Psi_m,\Omega^2(\phi)g_{\mu\nu}]\; ,\label{acao2}
\end{equation}
where $g_{\mu\nu}$ is a pure rank-2 tensor in the Einstein frame, and $R$ is 
the curvature scalar given by

\begin{equation}
R = R(\{\}) + 4\epsilon \alpha^2(\phi)\partial_{\mu} 
\phi \partial^{\mu} \phi \; . \label{R}
\end{equation}

It is interesting to call attention to the fact that the action in 
Eq.(\ref{acao2}) is obtained from Eq.(\ref{acao1}) by a conformal 
transformation of the kind 

\begin{equation}
\tilde{g}_{\mu\nu} = \Omega^2(\phi)g_{\mu\nu} \label{conform},
\end{equation}

\noindent
and by a redefinition of the quantity

\[G\Omega^2(\phi) = \tilde{\phi}^{-1}\; .\]

\noindent
This transformation makes it evident that any gravitational
phenomena will be affected by the variation of the gravitational 
{\it constant} $G$ in the scalar-tensor gravity, a feature that is 
exhibited through the definition of a new parameter

\[\alpha^2(\phi) \equiv \left( \frac{\partial \ln \Omega(\phi)}{\partial
\phi} \right)^2 = [2\omega(\tilde{\phi}) + 3]^{-1} ,\]

\noindent
which can be interpreted as the field-dependent coupling strength between 
matter and the scalar field. In order to turn our calculations as general 
as possible, we will not fix the factors $\Omega(\phi)$ and 
$\alpha (\phi)$, leaving them as arbitrary functions of the scalar field.

\section{Ordinary SCS plus a scalar field dynamics}

Let us now consider the dynamics of an ordinary SCS plus a scalar field, 
as the simple realization of our scenario for inducing optical 
activity in spacetime.  In the conformal frame, the Einstein equations are 
modified. A straightforward calculation shows that they turn into

\begin{eqnarray}
R_{\mu \nu }&=&2\xi \partial _\mu \phi \partial
_\nu \phi +8\pi G(T_{\mu \nu} - \frac{1}{2}g_{\mu
\nu}T) \; ,\label{escalar1}\\
G_{\mu\nu} & = & 2\xi \partial_{\mu}\phi\partial_{\nu}\phi -
\xi g_{\mu\nu}g^{\alpha\beta}\partial_{\alpha}\phi\partial_{\beta}
\phi + 8\pi G T_{\mu\nu} \; ,\label{eins}
\end{eqnarray}

\noindent
where $\xi $ is defined as

\begin{equation}
\xi(\phi) = 1- 2\epsilon \alpha^2(\phi)\; ,
\end{equation}

\noindent
which contains two contributions: one coming from the scalar-tensor term and 
another from the torsion.

In the scalar-tensor theory the Einstein equations are modified by
the presence of the field $\phi$ and are obtained by applying the
variational principle to Eq.(\ref{acao2}). Thus, the equation describing 
the dynamics of the field $\phi $ reads 

\begin{equation}
\Box _g\phi =-4\pi G\alpha (\phi )T  \label{eq1},
\end{equation}

\noindent
where
\begin{equation}
\Box_g \phi = \frac{\xi}{\sqrt{-g}}\partial_{\mu}
\left(\sqrt{-g} \partial^{\mu} \phi\right).
\end{equation}

Equation(\ref{eq1}) brings some  new information because it does not
appear in general relativity. It also shows us that a matter
distribution in the space behaves like a source for $\phi $, and,
as usual, for $g_{\mu \nu}$ as well.
Up to now we have dealt with the purely gravitational sector, in
what follows, however, we will introduce the action for the matter that
describes a cosmic string.

To describe the simplest cosmic string in a scalar-tensor theory, we require 
the matter dynamics  to be figured out from a complex scalar and a gauge
field, in an Abelian Higgs model with symmetry $U(1)$ whose action is  
given by

\begin{equation}
I_m = \int d^4x \sqrt{{\tilde g}}\left[-\frac{1}{2}D_{\mu}\Phi
(D^{\mu}\Phi)^*  - \frac{1}{4}F_{\mu \nu}F^{\mu \nu}  - V(|\Phi|)\right],
\end{equation}

\noindent
where $D_{\mu} \Phi = ( \partial_{\mu} + i qA_{\mu})\Phi $ is the covariant
derivative, whilst $V(|\Phi |)$ is the potential. The reason why the gauge 
fields do not minimally couple to torsion is well discussed in 
Refs.\cite{Hehl,Gaspperini}.  The field strengths are defined as usually: 
$ F_{\mu \nu} = \partial_{\mu}A_{\nu} - \partial_{\nu} A_{\mu}$, with $A_\mu$ 
being the gauge field. The action given by Eq.(\ref{acao1}) has a $U(1)$ 
symmetry, where the $U(1) $ group associated with the $\Phi$-field is 
broken by the vacuum and gives rise to vortices of the Nielsen-Olesen 
type\cite{Nielsen} (here written in terms of $(t,r,\theta,z)$ the usual 
cylindrical coordinates)

\begin{equation} 
\begin{array}{ll} \Phi = \varphi(r )e^{i\theta} \; ,\\
A_{\mu} = \frac{1}{q}[P(r) - 1]\delta^{\theta}_{\mu} \; .
\end{array}\label{vortex1} 
\end{equation}

The boundary conditions for the fields $\varphi(r) $ and $P(r)$ are the 
same as those of ordinary cosmic strings\cite{Nielsen}, namely

\begin{equation}
\begin{array}{ll}
\begin{array}{ll}
\varphi(r) = \eta, & r \rightarrow
\infty, \\
\varphi(r) =0, & r = 0, \end{array}&
\begin{array}{ll}
P(r) =0, & r \rightarrow \infty, \\
P(r) =1, & r= 0.  \end{array}
\end{array} \label{config1}
\end{equation}

The potential $V(\varphi )$ triggering the spontaneous
symmetry breaking can be built as 

\begin{equation} V(\varphi) = \frac{\lambda_{\varphi}}
{4} (\varphi ^2 - \eta^2)^2  , \end{equation}

\vspace{.5 true cm}

\noindent
where $\lambda_{\varphi}$ is a coupling constant. Constructed in this way, 
this potential possesses all the ingredients that makes it viable to drive 
the transition leading to the formation of a  cosmic string, as it is well 
stablished. 

If we solve the Einstein-Cartan equations and transform to
the Jordan-Fierz frame by using the conformal transformation of 
Eq.(\ref{conform}), we find that the metric of a static, straight axially 
symmetric SCS in scalar-tensor gravity, is given as\cite{Valdir}

\begin{equation}
ds^2 = [1+ 8G_0 \mu \xi ^{-1}\alpha^2(\phi_0) ln\rho/ r_0 ] 
[ -dt^2 + dz^2 + d\rho^2 + (1- 8G_0\mu)\rho^2 d\theta^2],\label{metric}
\end{equation}

\noindent
where $G_0$ is defined as $G_0 \equiv G \Omega^2(\phi_0)$ and we have used
the fact that for a cosmic string the linearized solution of 
Eq.(\ref{eq1}) is given by

\begin{equation}
\phi_{(1)} = 4 G_0
\alpha (\phi_0)\xi^{-1} \mu \ln \frac{\rho}{r_0}\label{phi} \; .
\end{equation}

The constant $r_0$ appearing in the Eqs.(\ref{metric}) and (\ref{phi})
is an integration constant, and is chosen, for convenience, as having the 
same order of magnitude as that of the string radius. The metric given 
by Eq.(\ref{metric}) can be obtained from the one corresponding to the 
superconducting cosmic string\cite{Valdir}, by considering the limit when 
the string current vanishes. 

\section{Chern-Simons-like coupling in a scalar-tensor SCS background }

Recent mesurements of optical polarization of light from quasars and 
galaxies  provide evidence that in some regions on 
the sky the radiation polarization vectors  are not randomly oriented 
as naturally expected, but rather they appear concentrated around a 
preferential direction\cite{HL01}. These are fundamental observations that, 
despite the sky coverage was a bit incomplete, may hint at the spacetime as 
exhibiting novel properties such as the recently claimed optical activity or 
birefringence\cite{das00,NR97,jain99}. An interesting set of potential 
explanations of this effect has been put forward in Refs.
\cite{Novikov,kuhne97,DM97,capo99,Sengupta}. Here we provide an alternative 
explanation of this phenomenon by considering the modification of the 
Maxwell action density that adds to it a Chern-Simons-like term

\begin{equation}
I_{eff} = \int d^4x\sqrt{\tilde g} \left(-\frac{1}{4} F_{\mu \nu}F^{\mu \nu} - 
\frac{1}{3!} \lambda \varepsilon^{\mu \nu \alpha \beta}  
F_{\mu \nu} A_\alpha S_\beta \right)\label{biref},
\end{equation}

\noindent
in which we couple the electromagnetic field $ F_{\mu \nu}$ 
and the vector potential $A_\alpha$ to the torsion vector $S_\beta$ 
(as we identify it here), responsible for the appearance of the preferred 
cosmic direction, as  suggested by the observations\cite{NR97}. The 
parameter $\lambda$ is the coupling constant of the theory, whose likely 
value will be estimated later upon current astronomical data sets. Next, 
we will investigate the r\^ole played by the Chern-Simons term in the 
scalar-tensor screwed cosmic string background.

In this paper we will consider the case where $S_{\mu}$ is a gradient of 
some scalar field  $\phi$ that preserves the gauge invariance but not the 
Lorentz invariance. In the scalar-tensor screwed cosmic string background we 
can consider this scalar $\phi$ as the dilaton. Thus, by using 
Eq.(\ref{torsion}) the torsion vector (in the Jordan-Fierz frame) is 
defined as 

\begin{equation}
S_{\mu} = \frac{3}{2} \partial_{\mu}\ln{\tilde \phi}\label{extra} \; .
\end{equation}

Substituting the linearized solution given by Eq.(\ref{phi}) 
into Eq.(\ref{extra}),
we have

\begin{equation}
S_{\mu} = - 3\alpha(\phi_0) \partial_{\mu}\phi_{(1)}\label{extra1}.
\end{equation}

In this context, the field equation of the eletromagnetic field becomes

\begin{equation}
\frac{1}{\sqrt{-\tilde g}} \partial_{\mu}\left[\sqrt{-\tilde g} 
F^{\mu \nu}\right]= \lambda \alpha(\phi_0)\tilde 
F^{\mu \nu}\partial_{\mu}\phi_{(1)}.\label{eletro}
\end{equation}

The solutions of the Eq.(\ref{eletro}) give us the corresponding 
dispersion relation

\begin{equation}
(k^{\alpha}k_{\alpha})^2 + (k^{\alpha}k_{\alpha})(S^{\beta}S_{\beta}) =
(k^{\alpha}S_{\alpha})^2 \label{dispersion} \; ,
\end{equation}

\noindent
with $\omega $ and ${\bf k}$ being the wave frequency and  wave vector, 
respectively, and form the 4-vector $k^{\alpha} = (\omega, \bf k)$; 
$k =|\bf k|$. Because of the magnitude of the effect observed, we do 
expect $S_{\alpha}$ to be small (see arguments and references in the next 
section) and  thence we can expand the dispersion relation 
Eq.(\ref{dispersion}) in powers of $S_{\alpha}$. By using 
Eq.(\ref{extra1}) and substituting $\phi_{(1)}$ as 
given by Eq.(\ref{phi}), we get, to first order, the following result

\begin{equation}
k_\pm = \omega \pm 2\lambda \xi^{-1} G_0 \mu \alpha^2(\phi_0) \hat{s} 
\cos({\gamma}),
\end{equation}

\noindent
where $\gamma$ is the angle between the propagation wavevector $\bf k$ of 
the radiation and the unit vector $\hat{s}$. We shall discuss later the 
relevant r\^ole of this wavevector components in face of the plane 
polarization of far out radio-galaxies and quasars. But before so doing 
let us generalize first this result to the case of a superconducting cosmic 
string.

\section{Screwed superconducting cosmic string effects  }

We have already studied the screwed superconducting cosmic string in a 
recent work \cite{Valdir}, from which further details can be obtained. 
In order to describe the simplest superconducting 
cosmic string in a scalar-tensor theory, we demand a more general matter 
dynamics to be constructed upon a couple of complex scalar and gauge fields, 
in an Abelian Higgs model whose action is written as 

\begin{equation}
I_m = \int d^4x \sqrt{{\tilde g}}\left[-\frac{1}{2}D_{\mu}\Phi
(D^{\mu}\Phi)^* - \frac{1}{2}D_{\mu} \Sigma (D^{\mu}
\Sigma)^* - \frac{1}{4}F_{\mu \nu}F^{\mu \nu} - \frac{1}{4}H_{\mu
\nu}H^{\mu \nu} - V(|\Phi|, |\Sigma |)\right],\label{acao3}
\end{equation}

\noindent
where $D_{\mu} \Sigma=(\partial_{\mu}+ iA_{\mu})\Sigma$ and $D_{\mu} \Phi=
(\partial_{\mu} + iC_{\mu})\Phi$ are the covariant derivatives. The field
strengths are defined as in the standard fashion: 
$F_{\mu \nu}=\partial_{\mu}A_{\nu}
-\partial_{\nu}A_{\mu}$ and $H_{\mu \nu}=\partial_{\mu}C_{\nu}
-\partial_{\nu}C_{\mu}$, with $A_{\mu}$ and $C_{\mu}$ being the gauge fields.
 
The action given by Eq.(\ref{acao3}) has a $U(1) \times U(1)'$ symmetry, 
where the $U(1) $ group associated with the $\varphi$-field is broken by 
the vacuum and gives rise to vortices of the Nielsen-Olesen 
type\cite{Nielsen}. The other $U(1)'$ symmetry, that we associate with the 
electromagnetism, acts on the  $\Sigma $-field. This symmetry is not broken 
by the vacuum. It is only broken in the interior of the defect. 
The $\Sigma$-field in the string core, where it acquires an expectation 
value, is responsible for a bosonic current being carried by the gauge 
field $A_{\mu}$. The only non-vanishing components of the gauge fields 
are $A_z(r)$ and $A_t(r)$, and the current-carrier phase may be expressed 
as $\zeta(z,t) = \omega_1 t - \omega_2z$ (see Ref.\cite{Valdir} for 
definitions). Notwithstanding, we focus only on the magnetic case. Their 
configurations are defined as

\begin{equation}
\begin{array}{ll}
\Sigma = \sigma(r)e^{i\zeta(z,t)},\\
A_{\mu} = \frac{1}{e}[A(r) - \frac{\partial \zeta(z,t)}{\partial
z}]\delta_{\mu}^{z},
\end{array}
\label{vortex2}
\end{equation}

\noindent
because of the rotational symmetry of the string itself. The fields
responsible for the cosmic string superconductivity satisfy 
the following boundary conditions

\begin{equation}
\begin{array}{ll}
\begin{array}{ll}
\frac{d}{d r}\sigma(r) = 0, &r=0, \\
\sigma(r) = 0, & r \rightarrow \infty, \end{array} &
\begin{array}{ll}
A(r) \neq 0, & r \rightarrow
\infty, \\
A(r) = 1, & r = 0.
\end{array}
\end{array}
\label{config2}
\end{equation}

The potential $V(\varphi, \sigma)$ triggering the spontaneous
symmetry breaking can be built in the most general case as

\begin{equation} V(\varphi, \sigma) = \frac{\lambda_{\varphi}}{4} (
\varphi ^2 - \eta^2)^2 + f_{\varphi \sigma}\varphi ^2\sigma ^2 +
\frac{\lambda_{\sigma}}{4}\sigma ^4 -
\frac{m_{\sigma}^2}{2}\sigma^2 , \end{equation}

\vspace{.5 true cm}

\noindent
where $\lambda_{\varphi}$, $\lambda_{\sigma}$, $f_{\varphi \sigma}$ and 
$m_{\sigma}$ are coupling constants. Constructed in this way, this potential 
possesses also all the ingredients so as to drive the formation of a 
superconducting cosmic string, as in analogy with the ordinary cosmic 
string case. 

In order to investigate the dynamics resulting from the interaction of 
electromagnetic radiation from quasars and radio-galaxies with the torsion 
field coupled to the spacetime of background generated by the SCS in this 
scalar-tensor theory, as encripted by the coupling term type Chern-Simons 
in Eq.(\ref{biref}),  we shall adopt the same procedure as in the last 
section to obtain the dilaton solution in this the superconducting case. 
For a superconducting cosmic string\cite{Valdir}, the dynamics defined by 
Eq.(\ref{eletro}) is complicated in itself, but if we assume that we are 
very far from the source then the gravitational coupling can be neglected, 
and so we are left with

\begin{equation}
\phi_{(1)} = 4 G_0
\alpha (\phi_0)\xi^{-1} (\mu + \tau - 
I^2) \ln \frac{\rho}{r_0}\label{phi1} \; .
\end{equation}

Working out the dynamics of the interaction pictured by Eq.(\ref{biref}), 
we get a dispersion relation. Once again, since we expect $S_{\alpha}$ to 
be small (see sound arguments in Ref.\cite{MSS02}), we can expand the 
dispersion relation Eq.(\ref{dispersion}) in powers of $S_{\alpha}$ 
to obtain, to first order,

\begin{equation}
k_\pm = \omega \pm 2\lambda \xi^{-1} G_0 (\mu + \tau - I^2) 
\alpha^2(\phi_0) \hat{s} \cos({\gamma})\label{k} \; .
\end{equation}

\noindent
In this case, the parameter I is the current in the
vortex and $\tau $ is the tension of the string.

We have already commented that the polarization of the radiation moving 
through an intervening magnetized intergalactic plasma must be removed 
from the measurements of the rotation of its polarization plane by using
the fact that Faraday rotation is proportional to the square of the wavelength.
In the case of the superconducting cosmic string in general relativity
there is no residual polarization related to the anisotropy after removing 
the Faraday rotation. In this paper we show that in scalar-tensor 
theories of gravity, this is not the situation. In fact, the polarization 
effect is present even after the Faraday rotation is removed.
The scalar coupling has a current carrying contribution given by $\tau $ 
and the current $I$ that appear in Eq.(\ref{k}). 

As for the comparison with the ordinary string as the background,  we note 
that the current $I$ has a strong negative contribution to the wavevector 
magnitude. This can be interpreted as if the photon has gained now a smaller 
wavenumber and consequently larger wavelength, and thus larger distances 
need to be traveled in order for the optical activity of spacetime to 
manifest on cosmic scales. In this case, these our results go on the same 
lines of those in the papers by Das, Jain and Mukherji\cite{das00} and by 
Kar et al., \cite{kar00}.

In the next section we will connect these our theoretical results with the 
conclusions drawn from data analisys of electromagnetic emission from 
radio-galaxies and quasars by Nodland and Ralston (NR97)\cite{NR97}; Jain 
and Ralston (JR99)\cite{jain99}; Das, Jain and Mukherji (DJM00)\cite{das00}; 
and more recently from the study of optical polarization of radio-emitting 
galaxies and quasars by Hutsem\'ekers and Lamy \cite{HL01}. Note that these 
new results by Hutsem\'ekers and Lamy \cite{HL01} appears to confirm their 
preliminary conclusions based on an earlier survey, and those in 
Refs.\cite{das00,NR97,jain99}, in which they analyzed measurements of the 
optical polarization properties of a limited sample of broad absortion-line 
radio-loud and radio-intermediate high-redshift quasars\cite{HL00}.

\section{Discussion}

Let us contextualize our theoretical results in the framework of the 
conclusions drawn from the analysis of observational data of quasar emission 
performed by NR97\cite{NR97}, JR99\cite{jain99} and DJM00\cite{das00}. In 
the analysis of the data of high-redshift radio-emitting galaxies and 
quasars: about 73 in NR97, 277 in JR99, and 231 in DJM00, they found 
correlations between the direction on the sky and the distance to a galaxy. 
The angle $\beta$ between the polarization vector and the galaxy's major 
axis is defined as 

\begin{equation}
<\beta> = \frac{1}{2}\frac{r}{\Lambda_s} \cos(\vec{k},\vec{s})\label{nr97},
\end{equation}

\noindent
where $< \beta >$ represents the mean rotation angle after Faraday's rotation 
is removed, $r$ is the distance to the galaxy, $\vec{k}$ the wavevector of the 
radiation, and $\vec{s}$ a unit vector defined by the direction on the 
sky: $\equiv (315^0 \pm 30^0,0^0 \pm 20^0)$, given here in equatorial 
celestial coordinates (r.a.,dec.).

The rotation of the polarization plane is a consequence of the difference 
in the propagation speed of the two modes $\kappa_+$, $\kappa_-$, the main 
dynamical quantitites computed above. This difference, defined as the 
angular gradient with respect to the radial (coordinate) distance, is 
expressed as 

\begin{equation}
\frac{1}{2}(\kappa_+ - \kappa_-) = \frac{d\beta}{dr}\; ,
\end{equation}

\noindent
where $\beta$ measures the specific entire rotation of the polarization 
plane, per unit length $r$, and is given once again by 
$\beta = \frac{1}{2}  \Lambda_s^{-1}  r \cos \gamma \;. $ In the case of the 
screwed cosmic string, the constant $\Lambda_s$, that encompasses the cosmic 
distance scale for the optical activity to be observed, can be written as a 
function of the string energy density $\mu $ as

\begin{equation}
\Lambda_s^{-1}= 4 G_0 \mu \lambda \xi ^{-1} {\Large \alpha }^2(\phi_0) \; .
\end{equation}

It is illustrative to consider a particular form for the arbitrary 
function $\alpha(\phi)$, corresponding to the Brans-Dicke theory, namely, 
$\Omega=e^{\alpha \phi}$, with $\alpha^2 = \frac{1}{2w +3}$, ($w$=cte).
Here we use the values for the scalar parameter $w > 2500$ such that the 
theory keeps consistent with solar system experiments made by using Very 
Large Baseline Interferometry
(VLBI) \cite{Eubanks}. In this case $G_0 = G_* \Omega^2(\phi_0) 
= \left(\frac{2\omega +3}{2\omega +4}\right) G_{eff}$ \cite{Dicke}, 
with $G_{eff} $ the Newtonian constant. In this way, for an ordinary SCS 
we have the following set of constraints

$$
\left|\begin{array}{ll}
G_0 \mu \sim 10^{-6} & \mbox{ from COBE data}\\
\alpha^2 \sim 2\times 10^{-2}
&\mbox{Brans-Dicke gravity}\\
\xi = 1 & \mbox{in our days}\\
\Lambda_s^{-1} = 10^{-32} {\rm eV}  & \mbox{ from Nodland and Ralston}
\end{array}\right|
$$

With these observational constraints one can put a limit in the value of 
the coupling constant of the theory, which yields: 
$\lambda \sim 10^{-26}$~eV. 

This result is very interesting because it shows that the optical effect 
could indeed be large in a theory that has a screwed cosmic string as the 
background in a scalar-tensor gravity, compared to the one expected from a 
theory that has not it, as for instance in 
Refs.\cite{kar00,kuhne97,DM97,MSS02}. In particular, the paper by 
Mukhopadhyaya, Sen and SenGupta\cite{MSS02} shows that a 
very large suppression of the torsion 
zero-mode on the visible brane occurs in their theory, and that that 
suppression scales down with the Planck mass M$_P = (\hbar c/G_N)^{1/2} 
\sim 10^{19}$~GeV. This means that their coupling constant 
is $\sim 10^{-28}$~eV. This suppression scale is two orders of magnitude 
smaller than the strength of coupling  attained in the present theory.  
In this way we show that the general scalar-tensor theory in the SCS 
background with torsion can give an interesting explanation to these 
observational data regarding the occurrence of optical activity in the 
spacetime spanning over cosmological distances. And more interesting yet, 
the result attained shows that the torsion field effect is more stronger 
than the one contemporary wisdom believed, and that it may indeed be being 
detected through this spacetime birefringence and optical activity 
phenomenology. This is a new view point as opposite to the one presented 
in Ref.\cite{MSS02}, where the {\it illusion} of a torsion-free universe 
appears to be supported in the framework of braneworld physics.

If both, the cosmological anisotropy that Nodland and Ralston\cite{NR97}; 
Jain and Ralston\cite{jain99}; and Das et al. (DJM00) \cite{das00} have 
claimed to exist in the direction (NR97) $\hat{s}  = 
([21 \pm 2]hrs,0^0 \pm 20^0)$ (r.a.,dec.), and the optical activity of 
spacetime evidenced in the  Hutsem\'ekers and Lamy \cite{HL01} surveys of 
quasar optical polarization vectors, are real, then they  may reconcile 
these modern observations with earlier 
ones\cite{birch82,gregory81,bracewell}. Although this birefringence 
phenomenon was contentious\cite{Eisenstein,Carroll,leahy,MP95}, 
at the moment its occurrence 
is vastly demonstrated \cite{HL00,HL01,das00,NR97,jain99}. Indeed, 
JR99\cite{jain99} and DJM00\cite{das00} suggested that the effect, which is 
clearly visible for the two highest redshifts in the DJM00 data set, would 
be more decisively demonstrated, as already pointed out by Carroll and 
Field\cite{Carroll}, when new data from very distant galaxies are 
incorporated to the sample. Thus, there seems to be room for viable 
explanations of the origin of such a preferred direction on the 
sky\cite{jain99,kuhne97}. 

\section{Closing remarks}

In the present paper we consider the possibility of explaining the quoted 
polarization effects in the framework of scalar-tensor gravity theories 
that couples to torsion. In our approach a screwed cosmic string, defining 
the background spacetime on which the interaction between the 
electromagnetic and torsion fields may take place, is able to create a 
propagating effect which may manifest outside the string itself. In this 
sense, being the  effect or phenomenon a real one, this model may also 
provide a consistent description of the NR97, JR99 and DJM00 conclusions by 
endowing the cosmic string spacetime with torsion. In fact, a possible 
connection of this effect with the existence of a cosmic rotation axis was 
pointed out in\cite{kuhne97,bracewell}, who suggested the discovery of a 
cosmic axis in NR97, and claimed that the observed phenomenon may be a 
realization of an ancient theoretical idea already presented in  
G\"odel's cosmology. An alternative mechanism was envisioned by Dobado and 
Maroto who proposed that the effect, if real, may be interpreted through 
the coupling of the electromagnetic field with a background torsion field 
created by charged fermions\cite{DM97}. Although a viable proposal, this 
suggestion exhibits a fundamental drawback: it is implicit that the sea of 
fermions is {\it isotropically distributed} in spacetime. Thence, the effect 
must be present in any other direction on the sky! Unfortunately, 
observations do not support this assumption. In addition, the model should 
also be put forward a bit further to cope with additional observational 
implications of the existence of such a chiral background of relic 
particles. This crucial constraint was overlooked in that paper. Meanwhile, 
the spacetime with torsion produced by the Kalb-Ramond field coupled 
gravitationally to the Maxwell field, in the heterotic string-inspired 
model of Ref.\cite{kar00} (and references therein), appears also to be a 
very sound proposal for explaining the origin of the optical activity 
observed in the synchroton radiation received from cosmic radio-galaxies 
and quasars. Once again, the mechanism suggests the just quoted isotropic 
effect, which not stands on the quoted observations evidencing the existence 
of a preferred direction on the sky. Moreover, there are several well-known 
open issues in the string theory used for that leaves the mechanism not 
fully solidly fundamented.

Put the other way round,  our theory is able to provide a consistent 
explanation for the occurrence of these phenomena regarding the curious 
propagation of electromagnetic waves over cosmic distances. Concommitantly, 
the theory can account also for the peculiar alignment of the spin axes 
observed in galaxies belonging to the supercluster 
Perseus-Piscis\cite{kuhne97,gregory81,bracewell}. If confirmed, this 
cosmic effect might prove an indication of the existence of a universal 
torsion (or shear) field and an universal spin, as well. 

We consider as a possible origin of such an effect the coupling 
between electromagnetic fields and some torsion field, as described by 
Eq.(\ref{biref}), in the background of a screwed cosmic string in 
scalar-tensor gravity theories, with no fermion fields. The use of 
alternative gravity theories is justified by the fact that Einstein's 
gravitation does not admit this  sort of phenomenology, and consequentely 
cannot provide any explanation to this puzzle. In this approach the screwed 
cosmic string background can be source of optical activity of spacetime even 
if the string has no spin. In a screwed cosmic string background the medium 
can be birefringent too due to the appearance of a torsion-dependent 
{\it deficit angle}, and this feature may induce an interaction of the 
eletromagnetic radiation with a scalar gradient by means of a Chern-Simons 
coupling. In this way, we can explain the optical activity without invoking 
an universal sea of either fermionic or axion-like matter pervading the 
spacetime as its source, as considered by other authors. At the same time, 
the fact that this theory is concommitant with several experiments and 
astronomical observations may hint at a gravitation theory beyond Einstein's 
general relativity.

\vspace{1 true cm}

{\large \bf Acknowledgments:} We would like to express our deep gratitude to
Prof. J.A. Helay\"el-Neto and Prof. R.W. K\"uhne for helpful suggestions on 
the subject of this paper. The authors would like to thank (CNPq-Brazil) for 
financial support. C. N. Ferreira also thank Centro Brasileiro de Pesquisas 
F\'{\i}sicas (CBPF) for hospitality. HJMC acknowledges support from a 
Grant-in-Aid of FAPERJ, the Science Foundation of Rio de Janeiro 
State (Brazil).

\end{document}